\documentclass[proceedings]{JHEP3}

\usepackage{epsfig}
\PrHEP{PrHEP hep2001}                   % Do not change this one
\conference{International Europhysics Conference on HEP}                % Do not change this one

\title{Theoretical developments on \\
       Quartic Gauge Boson Couplings at 
       LEP\thanks{Based on work done in collaboration with G.~Montagna, 
       M.~Moretti and O.~Nicrosini.}}

\author{\speaker{Fulvio Piccinini}                       % One and only one
        \thanks{On leave of absence from INFN - Sezione di Pavia.}\\  % author must be
        Theory Division, CERN, CH-1211, Geneva 23, Switzerland \\                                % inserted as
        E-mail: \email{Fulvio.Piccinini@cern.ch}}              % speaker

\author{Marco Osmo \\
        Dipartimento di Fisica Nucleare e Teorica - Universit\`a di Pavia, 
       and \\
       INFN - Sezione di Pavia, Via A. Bassi 6, Pavia, Italy\\
        E-mail: \email{Marco.Osmo@pv.infn.it}}              % speaker

\abstract{
The search for quartic anomalous gauge couplings (QAGC) at LEP requires 
appropriate predictions for the radiative processes 
$e^+ e^- \to \nu\bar\nu \gamma\gamma$, $e^+ e^- \to q\bar{q}\gamma\gamma$ and 
$e^+ e^- \to 4$~fermions+$\gamma$. The current knowledge on dimension-six 
operators giving rise to QAGC is briefly reviewed, 
together with their implementation in event generators. 
The accuracy of calculations based on real approximations 
(used up to now for the LEP experimental analysis) is examined by 
comparing them with the available exact matrix element calculations.
}

\begin{document}

\leftline{\tt Preprint~numbers: CERN-TH/20001-302, FNT/T2001/22}

\section{Introduction}
\label{intro}
Despite the striking success of the Standard Model (SM) in accommodating
the precision data collected at high-energy colliders, important 
tests of the theory, such as the non-abelian nature of the gauge 
symmetry and the mechanism of electroweak symmetry breaking, are still
at a beginning stage. To this end, gauge-boson self interactions 
play a key role. At present, triple gauge couplings are being 
probed at LEP~\cite{leptri} and the Tevatron~\cite{tevatron}, 
while direct 
measurements of quartic couplings only very recently became available through 
the study of radiative events 
at LEP~\cite{l3quad}--\cite{mb}. Actually, 
events with one or two isolated, hard photons are analysed at LEP, 
to search for anomalies in the sector of quartic gauge-boson couplings.
Only vertices involving at least one photon can be constrained, since 
quadrilinear interactions containing four massive gauge bosons give 
rise to a final state of three massive gauge bosons and 
are therefore beyond the potential 
of LEP, because of the lack of phase space. The processes considered in the 
experimental analyses are $e^+ e^- \to W^+ W^- \gamma$, 
$e^+ e^- \to Z \gamma \gamma$ and 
$e^+ e^- \to  \nu \bar\nu \gamma \gamma$~\cite{l3quad}--\cite{mb}.
The $W^+ W^- \gamma$ signature, which yields a four-fermion plus gamma 
final state, is interesting in order to test $WW\gamma \gamma$ 
and $WWZ\gamma$ vertices. For the $Z \gamma \gamma$ events, the 
final states due to the hadronic decays of the $Z$ boson with two 
jets and two visible photons are selected to probe the purely anomalous
vertex $ZZ\gamma \gamma$, which is of particular interest, being absent from 
the SM at tree level. The final state with two neutrinos and 
two acoplanar photons allows a study of the quartic $WW\gamma\gamma$ 
and $ZZ\gamma\gamma$ interactions. 

In the light of these experimental analyses, the aim of the present 
contribution is to review the state of the art of exact SM 
calculations for the processes $e^+ e^- \to 4$~fermions ($4f$) + $\gamma$, 
$ e^+ e^- \to q \bar{q} \gamma \gamma$ and 
$e^+ e^- \to  \nu \bar\nu \gamma \gamma$, including the effects of quartic
anomalous gauge couplings (QAGC).

\section{Theoretical approach}
\label{sec:1}
The theoretical framework of interest is the formalism of electroweak 
chiral Lagrangians. In such a scenario, QAGC involving four massive 
gauge bosons emerge as operators of dimension four at 
next-to-leading order,  
while QAGC with at least one photon originate from 
six (or higher) -dimensional operators at next-to-next-to-leading order. 
They are said genuinely anomalous if they do not induce 
new trilinear gauge interactions. 

Anomalous $WW\gamma\gamma$ and $ZZ\gamma\gamma$ vertices 
were originally introduced in ref.~\cite{bb}. In this paper, the authors 
show that, by assuming $C$ and $P$ conservation and 
further imposing $U(1)_{em}$ gauge invariance and $SU(2)_c$ custodial 
symmetry,  two independent Lorentz structures 
contribute to $WW\gamma\gamma$ and $ZZ\gamma\gamma$ interactions 
according to the following Lagrangians 
\begin{eqnarray}
{\cal L}_{0} &=& - \frac{e^2}{16} \, { \frac{a_0}{\Lambda^2} } \,
F_{\mu\nu}F^{\mu\nu} \vec{W}^\alpha \cdot {\vec W}_{\alpha} \nonumber\\
{\cal L}_{c} &=& - \frac{e^2}{16} \, { \frac{a_c}{\Lambda^2} } \,
F_{\mu\alpha}F^{\mu\beta} \vec{W}^\alpha \cdot {\vec W}_{\beta} \, ,
\label{bb}
\end{eqnarray}
where $F_{\mu\nu}$ is the 
electromagnetic field tensor, and $\vec{W}$ is a $SU(2)$ triplet describing 
the $W$ and $Z$ physical fields, 
$\cos\theta_w$ being the cosine of the weak mixing angle. In eq.~(\ref{bb}) 
$a_0$ and $a_c$ are (dimensionless) anomalous couplings, divided 
by an energy scale $\Lambda$, which has the meaning of scale of new physics.
Generally speaking, $\Lambda$ is in principle unknown and model-dependent. 
However, the ratios $a_i/\Lambda^2$ entering the phenomenological Lagrangians
can be meaningfully extracted from the data in a model-independent way.

The anomalous $WWZ\gamma$ vertex has been 
analysed in ref.~\cite{bela}, showing that, under the assumption of $C$, 
$P$ and $U(1)_{em}$ invariance, five additional Lorentz structures with 
respect to eq.~(\ref{bb}) contribute. 
It is further demonstrated that, 
by embedding all the structures related to 
$WW\gamma\gamma$, $ZZ\gamma\gamma$ and $WWZ\gamma$ vertices 
in $SU(2)\times U(1)$ 
gauge-invariant and $SU(2)_c$ symmetric combinations, fourteen $C$- and $P$- 
conserving operators are allowed, with $k_j^i$ parameters, which parametrize 
the strength of anomalous couplings. 

By allowing the violation of at least one discrete symmetry, but 
retaining $U(1)_{em}$ invariance and global custodial $SU(2)_c$ symmetry, 
additional terms can be introduced in the Lagrangian. In particular, 
in the literature three different contributions have been considered: 
${\cal L}_{n}$~\cite{an}--\cite{swer}, which violates $C$ and $CP$, 
$\tilde{{\cal L}_{0}}$~\cite{ddrw,mroth}, which violates $P$ and $CP$, 
and $\tilde{{\cal L}_{n}}$~\cite{ddrw,mroth}, which violates both $C$ and $P$, 
thus conserving $CP$.

In the approach of ref.~\cite{qagclett}, 
all the different operatorial structures 
contributing to the vertices analysed at LEP have been implemented 
directly at the Lagrangian level in the Monte Carlo codes 
{\tt NUNUGPV}~\cite{nunu} 
and {\tt WRAP}~\cite{wrap} with arbitrary $a_i$ coefficients. 
By means of appropriate relations between the $a_i$ 
parameters, both parametrizations available 
in the literature for QAGC, namely the parametrization in terms of $a_0, a_c, a_n$ 
couplings and the one in terms of $k_j^i$ coefficients, can be obtained, 
as shown explicitly in ref.~\cite{qagclett}.

By means of the above event generators, theoretical predictions for the 
processes $e^+ e^- \to \nu \bar\nu \gamma \gamma$, 
$e^+ e^- \to q \bar q \gamma \gamma$ and $e^+ e^- \to 4f + \gamma$ can be 
obtained. As an option of {\tt WRAP}, predictions
for the inclusive final states $WW\gamma$ and $Z\gamma\gamma$ can also be 
obtained, especially in order to compare them with results existing in the 
literature treating the $W$ and $Z$ bosons in the on-shell approximation, 
as those of refs.~\cite{bela,swer}. Recently the code {\tt RacoonWW} 
has been upgraded to include QAGC for the set of $4f + \gamma$ 
processes~\cite{ddrw,mroth}. 

\section{Discussion}

A detailed numerical investigation of the potentialities of the 
above-mentioned processes in the search for QAGC at LEP can be found in 
refs.~\cite{ddrw,qagclett}. Here only the main results are summarized. 

The options of $WW\gamma$ and $Z\gamma\gamma$ real production 
allowed a comparison of the predictions given by the Monte Carlo {\tt WRAP} 
with the results already present in the literature. In particular, such 
a comparison showed a very satisfactory agreement with the numerical 
results of ref.~\cite{bela}, while discrepancies were found with 
the results published in ref.~\cite{swer}, 
for the dependence of the $WW\gamma$ and $Z\gamma\gamma$ cross sections
on $a_0$ and $a_c$ parameters. More precisely, an opposite sign is present 
for the relative effects of $a_0$ and $a_c$ on the $WW\gamma$ cross section, 
as noted in ref.~\cite{ddrw}, 
whereas  this is not the case for the $Z\gamma\gamma$ 
final state.

An important issue, which can be addressed with complete matrix element 
calculations, is the reliability of the calculations performed in the 
narrow-width approximation, used so far in the experimental search for QAGC. 

As far as the neutral QAGC sector is concerned, 
the cross sections obtained with complete calculations for the final states 
$q \bar{q} \gamma\gamma$ and $\nu\bar\nu \gamma\gamma$ have been compared 
in ref.~\cite{qagclett} with 
the $Z\gamma \gamma$ approximation as functions of the parameters 
$a_0$ and $a_c$ at the c.m.\ energy of 200 GeV. 
In the case of the $q \bar{q} \gamma\gamma$ 
channel, a cut on the invariant mass of the jet--jet system around the $Z$ mass 
(80~GeV $\leq M_{q\bar{q}} \leq 100$~GeV) has been imposed, while for the 
$\nu\bar\nu \gamma\gamma$ final state a cut on the recoil mass again around the 
$Z$ mass (80~GeV $\leq M_{recoil} \leq 120$~GeV) has been adopted, in order 
to perform a consistent comparison with the $Z\gamma\gamma$ approximation. 
It is worth noting that these selection criteria are also adopted 
by the real experimental analysis. This comparison allows 
the effects due to the $\gamma$--$Z$ interference in the 
$q \bar{q} \gamma\gamma$ channel and to $W$--$Z$ interference in the 
$\nu\bar\nu \gamma\gamma$ one to be quantified, 
as well as the effects of the off-shellness of the $Z$ boson, 
in the extraction of limits on the QAGC. 
The numerical investigation of ref.~\cite{qagclett} shows an 
agreement between the integrated cross sections 
at the per cent level as a function of $a_0$ and $a_c$ variations inside 
the currently allowed experimental constraints, thus 
illustrating the reliability of the $Z\gamma\gamma$ approximation in view of the 
expected experimental precision. 
This conclusion also holds true for the 
differential distributions mostly sensitive 
to QAGC and considered in the experimental studies.

A similar analysis has been performed in ref.~\cite{qagclett} 
for a $4f + \gamma$ final state 
($e^+ e^- \to \mu {\bar \nu}_\mu u \bar d \gamma$), 
as computed by means of the exact 
calculation of {\tt WRAP}, in comparison with the $WW\gamma$ 
approximation considered in the literature~\cite{bela,swer}. 
The sensitivity to the anomalous couplings $a_0,a_c,a_n$ has been 
studied for the $4f + \gamma$ final state according to two
different event selections: no cuts on the invariant masses of the decay 
products, and cuts on the invariant masses of the decay products around 
the $W$ mass, i.e. 75 GeV 
$ \leq M_{u \bar{d}, \mu \bar{\nu}_{\mu}} \leq $ 85 GeV, 
in order to disentangle, as much as possible, the contributing Feynman graphs 
with two final-state resonant $W$ bosons. 
The $WW\gamma$ approximation predicts a quite different sensitivity with 
respect to the complete $4f + \gamma$ calculation.

By considering variations of the anomalous couplings within the allowed 
experimental bounds, differences at the ten per cent level are registered 
between the  $WW\gamma$ approximation and the $4f + \gamma$ prediction, 
when invariant mass cuts are imposed in the $4f + \gamma$ calculation. 
Notice that,  if invariant mass cuts are not considered, 
the $4f + \gamma$ cross section grows up by a factor of 2 with respect 
to the cross section in the presence of cuts. Therefore, the 
$WW\gamma$ approximation should be employed with due caution in QAGC 
studies, especially if we take into account that exact $4f + \gamma$ 
generators, such as {\tt WRAP}~\cite{wrap} and {\tt RacoonWW}~\cite{ddrw}, 
incorporate the effects of QAGC and are at our disposal for such experimental 
studies.

As a last remark, it is worth mentioning that, in realistic experimental 
analyses, the effects of ISR (properly simulated by means of the 
available event generators) should be considered, since they tend to diminish
the sensitivity on the QAGC at the level of some per cent.

\section{Conclusions}

The search for QAGC in radiative events at LEP demands precise predictions 
for the processes $e^+ e^- \to \nu\bar\nu \gamma\gamma$, 
$e^+ e^- \to q\bar{q}\gamma\gamma$ and 
$e^+ e^- \to 4$~fermions+$\gamma$. To this end, the available exact 
calculations for such processes, including the contribution of QAGC 
(and related event generators), have been reviewed. 

After a brief description of the possible Lagrangians giving rise to 
QAGC in vertices involving at least one photon, and of their implementation 
in existing event generators, 
comparisons between exact calculations and approximate results 
existing in the literature have been discussed. It turns out that, for 
the $\nu\bar\nu \gamma\gamma$ and $q\bar{q}\gamma\gamma$ final states, 
the $Z\gamma\gamma$ approximation works well, the interference 
effects present in the complete calculations being confined at the 
per cent level,
if appropriate cuts around the $Z$-boson mass are required. 
The $\nu \bar\nu \gamma \gamma$ final state with appropriate cuts on the 
recoil mass can be successfully exploited to extract limits on neutral QAGC, 
as a complementary channel to the $q \bar{q} \gamma \gamma$ one. As far as 
the 4~fermions+$\gamma$ final states are concerned, significant differences are 
seen between the exact calculation and the $WW\gamma$ approximation, even 
in the presence of invariant mass cuts around the $W$-boson mass.   

\vskip 24pt
\noindent
The authors are grateful to G.~Montagna for his careful 
reading of the manuscript. F. P. thanks the Conveners of the Parallel 
Session ``Test of the Standard Model'' for their kind invitation.

\end{document}